\documentclass[doublecol]{epl2}

\title{Topological quantum phase transition and the Berry phase near the
Fermi surface in hole-doped quantum wells} \shorttitle{Topological
quantum phase transition and the Berry phase near the Fermi
surface}

\author{Bin Zhou\inst{1,2}, Chao-Xing Liu\inst{3} \and Shun-Qing Shen\inst{1}%
} \shortauthor{B. Zhou \etal}

\institute{
  \inst{1} Department of Physics, and Center for Theoretical and Computational
Physics, The University of Hong Kong, Pokfulam Road, Hong Kong, China\\
  \inst{2} Department of Physics, Hubei University, Wuhan 430062,
  China\\
  \inst{3} Department of Physics and Center for Advanced Study, Tsinghua
University, Beijing 100084, China}

\pacs{73.43.Nq}{Quantum phase transitions} \pacs{03.65.Vf}{Phases:
geometric; dynamic or topological} \pacs{72.25.Dc}{Spin polarized
transport in semiconductors}

\abstract{ We propose a topological quantum phase
transition for quantum states with different Berry phases in
hole-doped III-V semiconductor quantum wells with bulk and
structure inversion asymmetry. The Berry phase of the occupied
Bloch states can be characteristic of topological metallic states.
It is found that the adjustment of thickness of the quantum well
may cause a transition of Berry phase in two-dimensional hole gas.
Correspondingly, the jump of spin Hall conductivity accompanies
the change of the Berry phase. This property is robust against the
impurity potentials in the system. Experimental detection of this
topological quantum phase transition is discussed.}

\begin{document}

\maketitle

\section{Introduction}

Topological properties of electron bands or Bloch states are fundamentally
important in characterizing quantum transverse transport of electrons in
metals and semiconductors. Studies of quantum Hall effect reveal the
topological origin of quantum Hall conductivity and the existence of novel
quantum states of matter \cite{Hall-effect}. Thouless \textit{et al.} \cite%
{Thouless82prl} found that quantum Hall conductivity can be expressed in
terms of Chern-Simon number of electron bands. Renewed interests of
anomalous Hall effect leads to an interpretation of "anomalous velocity" in
the Karplus-Luttinger formula for anomalous Hall conductivity as integration
of Berry curvatures of occupied Bloch states, which gives a geometric
insight of intrinsic contribution in ferromagnetic metals or semiconductors
\cite{Jungwirth02PRL,Haldane04prl}. It was also noticed that Berry phase or
Chern-Simon number may have very close relation to the intrinsic and quantum
spin Hall effect \cite{Shen04PRB,Chang05prb,Sheng06prl}. Very recently,
Bernevig \textit{et al.} proposed a topological quantum phase transition of
topological insulators in HgTe quantum wells \cite{Bernevig06SCI}.

Berry phase is acquired by a quantum state upon being transported
adiabatically around a loop in the parameter space \cite{Berry84}. It
reflects topological properties of bulk quantum states. Spin-orbit coupling
in semiconductors mixes electron Bloch states in the $k$ space with spin
degree of freedom. In some two-dimensional (2D) systems the Berry phase is
well defined for some band structures near the Fermi surface such as the
system with Rashba or Dresselhaus spin-orbit coupling \cite{Shen04PRB}. In
this paper, we investigate quantum size effect of the Berry phase near the
Fermi surface of heavy holes in III-V semiconductor quantum wells with bulk
and structure inversion asymmetry, and propose a topological quantum phase
transition for topological metallic states with different Berry phases when
changing the thickness of the quantum well. The anomaly or discontinuity of
quantum transverse transport of electron can be characteristic of this
topological quantum phase transition. As examples we study the spin Hall
conductance of the systems, and find that the spin Hall conductivity has a
jump near the transition point. This property is robust against the impurity
scattering and expected to be observed with the current experimental
technique.

\section{Model}

Consider a [001]-grown 2D quantum well of hole-doped III-V semiconductors.
We start with the model Hamiltonian for the valence band near the $\Gamma $
point in the $k$ space \cite{BPZ84Book,Bulaev05PRL},%
\begin{equation}
H_{\mathrm{bulk}}=H_{L}+H_{D}+H_{R}.  \label{bulk}
\end{equation}%
$H_{L}$ is the Luttinger Hamiltonian \cite{Luttinger56pr}%
\begin{equation}
H_{L}=\left( \gamma _{1}+\frac{5}{2}\gamma _{2}\right) \frac{\hbar ^{2}k^{2}%
}{2m}-\frac{\gamma _{2}}{m}\hbar ^{2}\left( \mathbf{k}\cdot \mathbf{S}%
\right) ^{2},
\end{equation}%
where $\gamma _{1}$, $\gamma _{2}$ are the material parameters, $%
k^{2}=k_{x}^{2}+k_{y}^{2}+k_{z}^{2}$, $m$ is the free electron mass, and $%
\mathbf{S}=\left( S_{x},S_{y},S_{z}\right) $ are $4\times 4$ matrices
corresponding to spin $3/2$. $H_{D}$ is the Dresselhaus spin-orbit coupling
caused by the bulk inversion asymmetry (BIA) \cite{Dresselhaus55}
\begin{equation}
H_{D}=-\frac{\gamma }{\eta }\left[ k_{x}\left( k_{y}^{2}-k_{z}^{2}\right)
S_{x}+c.p.\right] ,
\end{equation}%
where \textit{c.p. }stands for cyclic permutation of all indices ($x$, $y$, $%
z$), $\gamma $ is due to bulk inversion asymmetry, $\eta =\Delta
_{so}/(E_{g}+\Delta _{so})$, $\Delta _{so}$ is the split-off gap energy, $%
E_{g}$ is the band gap energy. $H_{R}$ is the Rashba spin-orbit coupling
term arising from structure inversion asymmetry (SIA) due to an asymmetry
confining potential \cite{Rashba60}%
\begin{equation}
H_{R}=\alpha \left( \mathbf{k}\times \mathbf{S}\right) \cdot \mathbf{e}_{z}%
\mathbf{,}
\end{equation}%
where $\alpha $ is a material parameter \cite{Winkler00prb} and $\mathbf{e}%
_{z}$ is the growth direction of the quantum well.

For a 2D quantum well with finite thickness $d$, the first heavy- and
light-hole bands have approximate relations of $\left\langle
k_{z}\right\rangle =0$ and $\left\langle k_{z}^{2}\right\rangle \simeq
\left( \pi /d\right) ^{2}$. If the thickness of quantum well is thin enough
such that the heavy hole (HH) and light hole (LH) bands are well separated.
In this paper, we limit our discussion to the case that only the first HH
band is significantly occupied. By means of the projection perturbation
method \cite{Winkler03Book,Zhu94PRB}, the bulk Hamiltonian Eq. (\ref{bulk})
is projected into the space of heavy holes,
\begin{eqnarray}
H_{hh} &=&\frac{\hbar ^{2}k^{2}}{2m_{hh}}+\lambda _{1}k^{2}\left(
k_{-}\sigma _{+}+k_{+}\sigma _{-}\right)  \nonumber \\
&&+\lambda _{2}\left( k_{+}^{3}\sigma _{+}+k_{-}^{3}\sigma _{-}\right)
+i\lambda _{3}\left( k_{-}^{3}\sigma _{+}-k_{+}^{3}\sigma _{-}\right)
\nonumber \\
&&+i\lambda _{4}k^{2}\left( k_{+}\sigma _{+}-k_{-}\sigma _{-}\right) ,
\label{HH}
\end{eqnarray}%
where $\sigma _{\alpha }$ are the Pauli matrices, $\sigma _{\pm }=\left(
\sigma _{x}\pm i\sigma _{y}\right) /2$, $k_{\pm }=k_{x}\pm ik_{y}$,
\begin{equation}
\lambda _{1}=\frac{3\gamma }{4\eta }\left( 1-\frac{3m^{2}\alpha ^{2}}{4\hbar
^{4}\gamma _{2}^{2}\left\langle k_{z}^{2}\right\rangle }\right) ,\lambda
_{2}=\frac{3m^{2}\gamma ^{3}\left\langle k_{z}^{2}\right\rangle }{16\hbar
^{4}\gamma _{2}^{2}\eta ^{3}},  \label{SOC1}
\end{equation}%
\begin{equation}
\lambda _{3}=\frac{3\alpha }{4\left\langle k_{z}^{2}\right\rangle }\left( 1-%
\frac{m^{2}\alpha ^{2}}{4\hbar ^{4}\gamma _{2}^{2}\left\langle
k_{z}^{2}\right\rangle }\right) ,\lambda _{4}=\frac{9m^{2}\alpha \gamma ^{2}%
}{16\hbar ^{4}\gamma _{2}^{2}\eta ^{2}},  \label{SOC2}
\end{equation}%
and the effective HH mass
\begin{equation}
m_{hh}=m\left[ \gamma _{1}+\gamma _{2}-\frac{3m^{2}\left( \alpha ^{2}+\beta
^{2}+2\alpha \beta \sin 2\theta \right) }{4\hbar ^{4}\left\langle
k_{z}^{2}\right\rangle \gamma _{2}}\right] ^{-1},
\end{equation}%
with $\beta =\gamma \left\langle k_{z}^{2}\right\rangle /\eta $. The band
mixing between the light and heavy holes is taken into account as the
effective spin-orbit couplings. Correspondingly, the projected spin operator
$S_{z}$ has the form,
\begin{equation}
S_{hh}^{z}=\left[ \frac{3}{2}-\frac{3m^{2}\left( \alpha ^{2}+\beta
^{2}+2\alpha \beta \sin 2\theta \right) k^{2}}{16\hbar ^{4}\left\langle
k_{z}^{2}\right\rangle ^{2}\gamma _{2}^{2}}\right] \sigma _{z}.
\end{equation}%
As a result, there are four types of effective cubic spin-orbit coupling. $%
\lambda _{1}$, $\lambda _{2}$, and $\lambda _{3}$ can be adjusted by
thickness $d$ of quantum well through $\left\langle k_{z}^{2}\right\rangle $%
, and $\lambda _{4}$ is determined by the material parameters.

\section{The Berry phase}

Now we come to discuss topological properties of band structure and their
quantum-size effect$.$ The effective $2\times 2$ Hamiltonian (\ref{HH}) can
be diagonalized exactly in the $k$ space. The two eigenstates are%
\begin{equation}
\left\vert k,+\right\rangle =\frac{1}{\sqrt{2}}\left(
\begin{array}{c}
1 \\
e^{i\varphi }%
\end{array}%
\right) ,\left\vert k,-\right\rangle =\frac{1}{\sqrt{2}}\left(
\begin{array}{c}
e^{-i\varphi } \\
-1%
\end{array}%
\right) ,  \label{eigensates}
\end{equation}%
where $\varphi $ is given by%
\begin{equation}
\tan \varphi =\frac{\lambda _{1}\sin \theta -\lambda _{2}\sin 3\theta
-\lambda _{3}\cos 3\theta -\lambda _{4}\cos \theta }{\lambda _{1}\cos \theta
+\lambda _{2}\cos 3\theta +\lambda _{3}\sin 3\theta -\lambda _{4}\sin \theta
},
\end{equation}%
and $\tan \theta =k_{y}/k_{x}$.

\subsection{The case without SIA}

We first only consider the case of the pure BIA, i.e, $\alpha =0$. In this
case $\lambda _{3}=\lambda _{4}=0,$ and $\lambda _{1}=3\gamma /\left( 4\eta
\right) ,\lambda _{2}=3m^{2}\gamma ^{3}\left\langle k_{z}^{2}\right\rangle
/\left( 16\hbar ^{4}\gamma _{2}^{2}\eta ^{3}\right) .$ Thus, the two-band
effective Hamiltonian is reduced to
\begin{eqnarray}
H_{hh}^{\prime } &=&\frac{\hbar ^{2}k^{2}}{2m_{hh}}+\lambda _{1}k^{2}\left(
k_{-}\sigma _{+}+k_{+}\sigma _{-}\right)  \nonumber \\
&&+\lambda _{2}\left( k_{+}^{3}\sigma _{+}+k_{-}^{3}\sigma _{-}\right) ,
\label{HH1}
\end{eqnarray}%
where $m_{hh}=m\left[ \gamma _{1}+\gamma _{2}-3m^{2}\beta ^{2}/\left( 4\hbar
^{4}\left\langle k_{z}^{2}\right\rangle \gamma _{2}\right) \right] ^{-1}$. $%
\lambda _{1}$ is independent of the thickness $d$, but $\lambda _{2}$ is
proportional to $1/d^{2}$. There exists a critical thickness $d_{c1}$ such
that $\lambda _{1}=\lambda _{2}$. The value of the critical thickness $%
d_{c1}=m\pi \gamma /\left( 2\hbar ^{2}\gamma _{2}\eta \right) ,$ which is
determined by material-specific parameters. Table I gives material
parameters of some III-V semiconductors (after Refs. \cite%
{Miller03PRL,Winkler03Book}) and calculated critical thickness $d_{c1}$.

\begin{table}[tbp]
\caption{Material parameters of selected III-Vs and calculated critical
thickness $d_{c1}$.}
\label{tab.1}
\begin{center}
\begin{tabular}{llllll}
\hline\hline
\ \ \ \  & GaAs & \ InAs & GaSb & InSb \  & InP \  \\ \hline
$E_{g}$ (eV) & \multicolumn{1}{c}{$1.519$} & \multicolumn{1}{c}{$0.418$} &
\multicolumn{1}{c}{$0.813$} & \multicolumn{1}{c}{$0.237$} &
\multicolumn{1}{c}{$1.423$} \\
$\Delta _{so}$ (eV) & \multicolumn{1}{c}{$0.341$} & \multicolumn{1}{c}{$0.38$%
} & \multicolumn{1}{c}{$0.75$} & \multicolumn{1}{c}{$0.81$} &
\multicolumn{1}{c}{$0.110$} \\
$\gamma _{1}$ & \multicolumn{1}{c}{$6.85$} & \multicolumn{1}{c}{$20.4$} &
\multicolumn{1}{c}{$13.3$} & \multicolumn{1}{c}{$37.1$} & \multicolumn{1}{c}{%
$4.95$} \\
$\gamma _{2}$ & \multicolumn{1}{c}{$2.1$} & \multicolumn{1}{c}{$8.3$} &
\multicolumn{1}{c}{$4.4$} & \multicolumn{1}{c}{$16.5$} & \multicolumn{1}{c}{$%
1.65$} \\
$\gamma \mathrm{{\ (eV.\mathring{A}}^{3}}$) & \multicolumn{1}{c}{$28$} &
\multicolumn{1}{c}{$130$} & \multicolumn{1}{c}{$187$} & \multicolumn{1}{c}{$%
226.8$} & \multicolumn{1}{c}{$8.5$} \\
$d_{c1}$ (nm) & \multicolumn{1}{c}{$1.50$} & \multicolumn{1}{c}{$0.68$} &
\multicolumn{1}{c}{$1.83$} & \multicolumn{1}{c}{$0.37$} & \multicolumn{1}{c}{%
$1.48$} \\ \hline\hline
\end{tabular}%
\end{center}
\end{table}

The two dispersion relations corresponding to the eigenstates (\ref%
{eigensates}) are
\begin{equation}
E_{\mu }\left( k,\theta \right) =\frac{\hbar ^{2}k^{2}}{2m_{hh}}+\mu \lambda
(\theta )k^{3},  \label{dispersion}
\end{equation}%
where $\mu =\pm 1$ and $\lambda (\theta )=\sqrt{\lambda _{1}^{2}+\lambda
_{2}^{2}+2\lambda _{1}\lambda _{2}\cos 4\theta }$. In general the two bands
do not crossover except at $k=0$. In the case of $\lambda _{1}=\lambda _{2}$%
, \textit{i.e.}, at the critical point of $d=d_{c1}$, the two bands become
degenerate at $\theta =\pm \pi /4$ and $\pm 3\pi /4$. The Fermi surfaces and
dispersion relations along [110] axis are plotted in Fig. 1 for three cases
at or near the critical point of $\lambda _{1}=\lambda _{2}$. We note that
the validity of the above model is restricted to sufficiently small wave
numbers and hole densities, which is similar to the case of cubic Rashba
model \cite{Schliemann05PRB}.

\begin{figure}[tbp]
\includegraphics[width=7cm]{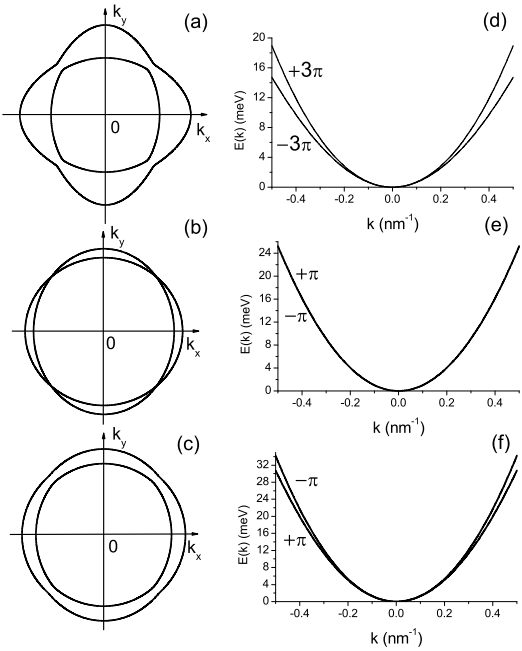}
\caption{{}Fermi surfaces and dispersion branches of heavy hole along [110]
direction for different thickness $d$ of GaAs quantum well. (a) fermi
surface for $d<$ $d_{c1}$; (b) fermi surface for $d=$ $d_{c1}$; (c) fermi
surface for $d>$ $d_{c1}$; (d) dispersion branches along [110] direction for
$d<$ $d_{c1}$; (e) dispersion branches along [110] direction for $d=$ $%
d_{c1} $; (f) dispersion branches along [110] direction for $d>$ $d_{c1}$.
The material parameters of GaAs are given in Table I. $\pm \protect\pi ,\pm 3%
\protect\pi $ stand for Berry phases.}
\end{figure}

The topological property of hole band is revealed by the vector potential
for the Berry phase in the $k$ space,%
\begin{eqnarray}
\mathbf{A}_{\mu } &=&i\left\langle k,\mu \left\vert \nabla _{\mathbf{k}%
}\right\vert k,\mu \right\rangle  \nonumber \\
&=&-\frac{\mu }{2k}\frac{\lambda _{1}^{2}-3\lambda _{2}^{2}-2\lambda
_{1}\lambda _{2}\cos 4\theta }{\lambda _{1}^{2}+\lambda _{2}^{2}+2\lambda
_{1}\lambda _{2}\cos 4\theta }\mathbf{e}_{\theta }.
\end{eqnarray}%
The associated Berry curvature is
\begin{equation}
\nabla _{\mathbf{k}}\times \mathbf{A}_{\mu }=\gamma _{\mu }\delta (\mathbf{k}%
)\mathbf{e}_{z},
\end{equation}%
where
\begin{equation}
\gamma _{\mu }=\mu \left[ \pi -2\pi \frac{\left( \lambda _{1}^{2}-\lambda
_{2}^{2}\right) }{\left\vert \lambda _{1}^{2}-\lambda _{2}^{2}\right\vert }%
\right]
\end{equation}%
for $\lambda _{1}\neq \lambda _{2}$ and $\mu \pi $ for $\lambda _{1}=\lambda
_{2}$. The phases are opposite for the two bands. The singularity at $%
\mathbf{k}=0$ indicates the existence of Berry phase flux or 2D magnetic
monopole in the $k$ space. We notice that the two types of spin-orbit
coupling in Eq. (\ref{HH1}) have quite different contributions to the Berry
phase. When the first term dominates $\lambda _{1}>$ $\lambda _{2}$, $\gamma
_{\mu }=-\mu \pi $ and oppositely $\gamma _{\mu }=\mu 3\pi $. At the
critical point of $\lambda _{1}=\lambda _{2}$, $\gamma _{\mu }=\mu \pi $.
According to the Stokes' theorem, $\gamma _{\mu }$ is exactly the Berry
phase \cite{Berry84,Note}, which is acquired by a state upon being
transported around an arbitrary loop $C$ including the origin of $\mathbf{k}%
=0$\ in the $k$ space, $\gamma _{\mu }=\oint_{C}d\mathbf{k}\cdot \mathbf{A}%
_{\mu }$\textbf{.} From these results, it indicates that adjustment of the
thickness $d$ near the critical point $d_{c1}$ may change the value of the $%
\lambda _{2}$, and further causes a change of Berry phase of $\gamma _{\mu
}=-\mu \pi $ to $\gamma _{\mu }=\mu 3\pi $ in the system or vice verse.
Since this Berry phase reflects the global topological properties of hole
bands in the $k$ space, it is believed that this phase transition is
topological.

\subsection{The case with BIA and SIA}

Now we will consider the system with both BIA and SIA. In the following, we
use material parameters of III-V semiconductor GaAs given in Table I, and
take $\alpha =0.01$ eV.nm. Variation of Berry phase $\gamma _{\mu }$ with
the thickness $d$ is plotted in Fig. 2. Due to SIA, a new step of the Berry
phase appears near $1.5$ nm. Furthermore, with the increase of the thickness
the Berry phase can transit from $\gamma _{\mu }=-\mu \pi $ to $\gamma _{\mu
}=-\mu 3\pi $ at $d=12.3$ nm.

\begin{figure}[tbp]
\includegraphics[width=8cm]{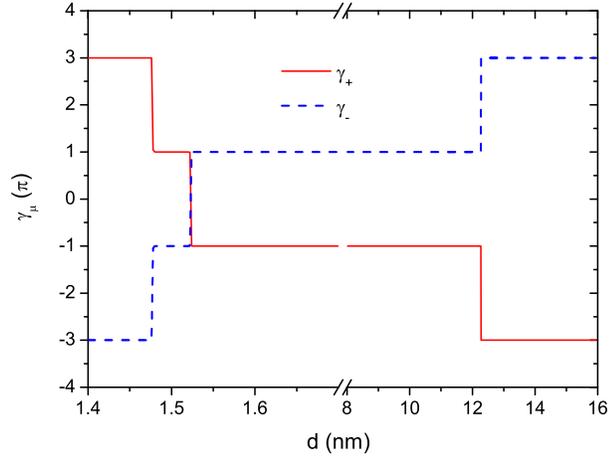}
\caption{{}Variation of Berry phase $\protect\gamma _{\protect\mu }$ with
the thickness $d$ of GaAs quantum well with BIA and SIA. The material
parameters are given in the text. The solid line (red) correspond to $%
\protect\gamma _{+}$; the dashed line (blue) to $\protect\gamma _{-}$.}
\end{figure}

Though there exist several transition points of the Berry phase, in the
following discussion, we focus on the regime near the transition at $%
d_{c2}=12.3$ nm. In this regime $\lambda _{1}$ and $\lambda _{3}$ are much
larger than $\lambda _{2}$ and $\lambda _{4}$. For simplification, we
neglect $\lambda _{2}$ and $\lambda _{4}$, and the effective Hamiltonian is
\begin{eqnarray}
\tilde{H}_{hh} &=&\frac{\hbar ^{2}k^{2}}{2m_{hh}}+\lambda _{1}k^{2}\left(
k_{-}\sigma _{+}+k_{+}\sigma _{-}\right)  \nonumber \\
&&+i\lambda _{3}\left( k_{-}^{3}\sigma _{+}-k_{+}^{3}\sigma _{-}\right) .
\label{HH2}
\end{eqnarray}%
The two dispersion relations have the same forms as Eq. (\ref{dispersion})
with $\lambda (\theta )=\sqrt{\lambda _{1}^{2}+\lambda _{3}^{2}+2\lambda
_{1}\lambda _{3}\sin 2\theta }$. In general the two bands do not crossover
except at $k=0$. In the case of $\lambda _{1}=\lambda _{3}$, or $d=d_{c2}$ ($%
d_{c2}$ shifts to $12.1$ nm due to the ignorance of $\lambda _{2}$ and $%
\lambda _{4}$ and remaining the definition of $\lambda _{1}$ and $\lambda
_{3}$ in Eqs. (\ref{SOC1}) and (\ref{SOC2})), the two bands become
degenerate at $\theta =3\pi /4$ and $7\pi /4$. The Fermi surfaces are
plotted in Fig. 3 at or near the critical point of $\lambda _{1}=\lambda
_{3} $.

\begin{figure}[tbp]
\includegraphics[width=8cm]{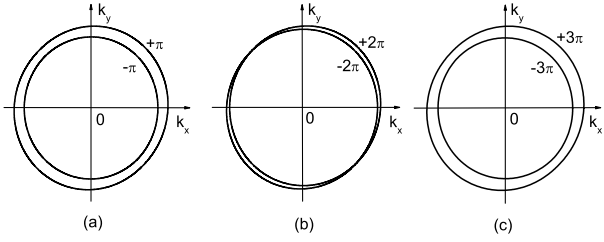}
\caption{{}Fermi surfaces for different thickness $d$ of GaAs quantum well.
(a) fermi surface for $d<d_{c2}$; (b) fermi surface for $d=$ $d_{c2}$; (b)
fermi surface for $d>$ $d_{c2}$. $\pm \protect\pi $, $\pm 2\protect\pi $,
and $\pm 3\protect\pi $ stand for Berry phase.}
\end{figure}

In this case the vector potential for the Berry phase in the $k$ space is
given%
\begin{equation}
\mathbf{A}_{\mu }=-\frac{\mu }{2k}\frac{\lambda _{1}^{2}+3\lambda
_{3}^{2}+4\lambda _{1}\lambda _{3}\sin 2\theta }{\lambda _{1}^{2}+\lambda
_{3}^{2}+2\lambda _{1}\lambda _{3}\sin 2\theta }\mathbf{e}_{\theta },
\end{equation}%
and thus the Berry phase is
\begin{equation}
\gamma _{\mu }=-\mu \left[ 2\pi -\pi \frac{\lambda _{1}^{2}-\lambda _{3}^{2}%
}{\left\vert \lambda _{1}^{2}-\lambda _{3}^{2}\right\vert }\right]
\end{equation}%
for $\lambda _{1}\neq \lambda _{3}$ and $-\mu 2\pi $ for $\lambda
_{1}=\lambda _{3}$. It follows that adjustment of the thickness $d$ may
cause a transition from the Berry phase of $\gamma _{\mu }=-\mu \pi $ to $%
\gamma _{\mu }=-\mu 3\pi $ in the system or vice verse.

On the other hand, we note that the strength of $\alpha $ is another
parameter which can be modified by a gate field. If the thickness $d$ of
quantum well is fixed, the change of $\alpha $ can also induce the change of
Berry phase. For example for a GaAs quantum well with $d=10$ nm, the
critical value $\alpha _{c}=0.014$ eV.nm at which the Berry phase can vary
from $\gamma _{\mu }=-\mu \pi $ to $\gamma _{\mu }=-\mu 3\pi $.

\section{The topological quantum phase transition and discontinuity of spin
Hall conductance}

The free electron gas described by the effective Hamiltonian is obviously
metallic. The spin-orbit coupling makes the electrons near the Fermi surface
to possess different topological properties in the $k$ space. The question
is whether these metallic states with different Berry phases are different
from each other such that the Berry phase can be characteristic of these
quantum metallic states. To reveal the relevant physical properties of these
metallic states, we study the spin Hall effect of this system, which has
attracted a lot of interests in recent years \cite{Murakami03sci,Sinova06ssc}%
. Without loss of generality, we shall focus on the effective Hamiltonian in
Eq. (\ref{HH2}) to explore the physical consequence of the change of the
Berry phase near $d_{c2}.$ The other two transition points of the Berry
phase require much thinner thickness.

For a realistic calculation we need to consider the effect of impurities,
which has drastic influence on some systems such as linear Rashba system
\cite{Inoue04prb,Sinova06ssc}. For simplicity, we consider $\tilde{H}_{hh}$
in Eq. (\ref{HH2}) with nonmagnetic impurities with short-ranged potential:
\begin{equation}
V\left( \mathbf{r}\right) =V_{0}\sum_{i}\delta \left( \mathbf{r}-\mathbf{R}%
_{i}\right) ,
\end{equation}%
where $V_{0}$ is the strength of impurities. The retarded Green function can
be written as
\begin{equation}
G^{R}(\mathbf{k},E,\mathbf{\Sigma }^{R})=(E-\tilde{H}_{hh}-\mathbf{\Sigma }%
^{R})^{-1},
\end{equation}%
where the self energy $\mathbf{\Sigma }^{R}$ is obtained in the Born
approximation by solving the self-consistent equation,
\begin{equation}
\mathbf{\Sigma }^{R}=n_{i}V_{0}^{2}\int \frac{d\mathbf{k}}{(2\pi )^{2}}G^{R}(%
\mathbf{k},E,\mathbf{\Sigma }^{R}),  \label{self}
\end{equation}%
where $n_{i}$ is the density of impurity. In this problem, the self-energy
has a diagonal form, $\mathbf{\Sigma }^{R}=\xi ^{R}\mathbf{I}$ with $\mathbf{%
I}$ being the $2\times 2$ unit matrix. The spin current operator $J_{y}^{z}$
is defined as $J_{y}^{z}=\left( \hbar /2\right) \{v_{y},S_{hh}^{z}\}$, and
the velocity operators are $v_{x}\equiv \lbrack x,\tilde{H}_{hh}]/(i\hbar )$
and $v_{y}\equiv \lbrack y,\tilde{H}_{hh}]/(i\hbar )$. To calculate the
linear response of spin current to the \textit{dc} electric field, we take
the vertex correction \cite{Inoue04prb}, and the spin Hall conductivity
reads
\begin{equation}
\sigma _{yx}^{z}=\frac{e\hbar }{2\pi }\int \frac{d\mathbf{k}}{\left( 2\pi
\right) ^{2}}\mathrm{Tr}_{\sigma }\left[ J_{y}^{z}G^{R}\mathbf{V}_{x}G^{A}%
\right] ,  \label{shc}
\end{equation}%
where $\mathbf{V}_{x}$ is the velocity operator with the vertex correction.
The self-consistent vertex equation includes the diagrams with impurity
ladders into the vertex part \cite{Abrikosov63Book}
\begin{equation}
\mathbf{V}_{x}=v_{x}+n_{i}V_{0}^{2}\int \frac{d\mathbf{k}}{\left( 2\pi
\right) ^{2}}G^{R}\mathbf{V}_{x}G^{A}.  \label{vertex}
\end{equation}%
The solution of $\mathbf{V}_{x}$ has the form $\mathbf{V}_{x}=v_{x}+%
\sum_{i}c_{i}\sigma _{i}$, and can be determined self-consistently. The
detailed calculation gives the solution $c_{z}=0$ and
\begin{equation}
c_{x}=\frac{A_{a}A_{d}+A_{b}A_{10}}{A_{c}A_{d}-A_{10}^{2}},
\end{equation}%
\begin{equation}
c_{y}=\frac{A_{a}A_{10}+A_{b}A_{c}}{A_{c}A_{d}-A_{10}^{2}},
\end{equation}%
where $A_{a}=A_{1}+A_{2}+A_{3}$, $A_{b}=A_{4}+A_{5}+A_{6}$, $%
A_{c}=1-A_{7}-A_{8}$, and $A_{d}=1-A_{7}-A_{9}$. The relevant parameters are%
\begin{equation}
A_{i}=\frac{n_{i}V_{0}^{2}}{4}\sum_{\mu ,\nu }\int \frac{d\mathbf{k}}{(2\pi
)^{2}}\Gamma _{i}^{\mu \nu }G_{\mu }^{R}G_{\nu }^{A},
\end{equation}%
where
\begin{equation}
G_{\mu }^{R(A)}=\frac{1}{E-E_{\mu }-\xi ^{R(A)}},
\end{equation}%
\begin{equation}
\Gamma _{1}^{\mu \nu }=\frac{(\mu +\nu )\kappa _{x}}{\kappa }\frac{\partial
\varepsilon }{\hbar \partial k_{x}},\Gamma _{2}^{\mu \nu }=(1-\mu \nu )\frac{%
\partial \kappa _{x}}{\hbar \partial k_{x}},
\end{equation}%
\begin{equation}
\Gamma _{3}^{\mu \nu }=\frac{\mu \nu \kappa _{x}}{\kappa }\frac{\partial
\kappa }{\hbar \partial k_{x}},\Gamma _{4}^{\mu \nu }=\frac{(\mu +\nu
)\kappa _{y}}{\kappa }\frac{\partial \varepsilon }{\hbar \partial k_{x}},
\end{equation}%
\begin{equation}
\Gamma _{5}^{\mu \nu }=(1-\mu \nu )\frac{\partial \kappa _{y}}{\hbar
\partial k_{x}},\Gamma _{6}^{\mu \nu }=\frac{\mu \nu \kappa _{y}}{\hbar
\kappa }\frac{\partial \kappa }{\partial k_{x}},
\end{equation}%
\begin{equation}
\Gamma _{7}^{\mu \nu }=1-\mu \nu ,\Gamma _{8}^{\mu \nu }=\frac{2\mu \nu
\kappa _{x}^{2}}{\kappa ^{2}},
\end{equation}%
\begin{equation}
\Gamma _{9}^{\mu \nu }=\frac{2\mu \nu \kappa _{y}^{2}}{\kappa ^{2}},\Gamma
_{10}^{\mu \nu }=\frac{2\mu \nu \kappa _{x}\kappa _{y}}{\kappa ^{2}},
\end{equation}%
with
\begin{equation}
\kappa _{x}=k_{x}k^{2}\lambda _{1}-k_{y}\left( k_{y}^{2}-3k_{x}^{2}\right)
\lambda _{3},
\end{equation}%
\begin{equation}
\kappa _{y}=k_{y}k^{2}\lambda _{1}-k_{x}\left( k_{x}^{2}-3k_{y}^{2}\right)
\lambda _{3},
\end{equation}%
$\kappa ^{2}=\kappa _{x}^{2}+\kappa _{y}^{2}$, and $\varepsilon =\hbar
^{2}k^{2}/\left( 2m_{hh}\right) $. Using the self-consistent solution of
self energies in Eq. (\ref{self}), we can calculate the spin Hall
conductivity explicitly. For numerical calculation here we adopt the
material parameters of GaAs given above and Fermi energy $E_{f}=2.5$ meV
which is close to the bottom of the bands.

Before doing numerical calculation, we first consider the problem in the
clean limit. The vertex-corrected velocity consists of two parts, the bare
velocity $v_{x}$ and vertex correction $\delta v_{x}=c_{x}\sigma
_{x}+c_{y}\sigma _{y}$. Correspondingly, the spin Hall conductivity in Eq. (%
\ref{shc}) can be divided into the intrinsic part and the vertex correction
part. Denote by $\tau ^{-1}=-\frac{2}{\hbar }$Im$(\xi ^{R})$ the life time.
In the clean limit of $n_{i}\rightarrow 0$, $\tau \rightarrow +\infty $, the
intrinsic part of spin Hall conductivity gives%
\begin{eqnarray}
\sigma _{yx}^{z,int} &=&\frac{3e\hbar ^{2}}{16\pi ^{2}}\int \frac{\sin
^{2}\theta d\theta }{m_{hh}\lambda ^{3}}\left( \frac{1}{k_{f}^{-}}-\frac{1}{%
k_{f}^{+}}\right)  \nonumber \\
&&\times \left( \lambda _{1}^{2}+3\lambda _{3}^{2}+4\lambda _{1}\lambda
_{3}\sin 2\theta \right) ,
\end{eqnarray}%
where $k_{f}^{\pm }(\theta )$ are $\theta $-dependent Fermi momenta of two
bands. This can be also obtained from the Kubo formula explicitly. In the
low density of carriers, $1/k_{f}^{-}-1/k_{f}^{+}\approx -2m_{hh}\lambda
(\theta )/\hbar ^{2}$. Using this formula we reach at an explicit relation
between the intrinsic part of the spin Hall conductivity and the Berry phase
near the Fermi surface
\begin{equation}
\sigma _{yx}^{z,int}=\frac{3e}{16\pi ^{2}}\sum_{\mu }\mu \gamma _{\mu }(d).
\end{equation}%
A similar relation has already been obtained for the system with the Rashba
and Dresselhaus spin-orbit coupling once the two conduction bands are
occupied simultaneously \cite{Shen04PRB,Chang05prb}. This relation reflects
the close relation between the spin Hall conductance and the topological
properties of the Fermi surface. Taking into account the vertex correction,
the total spin Hall conductivity in the clean limit is
\begin{equation}
\sigma _{yx}^{z}=-\frac{3e}{8\pi }\left[ 1-\frac{\hbar }{k_{f}^{2}\lambda
_{1}}(c_{x}+\frac{\lambda _{3}}{\lambda _{1}}c_{y})\right]
\end{equation}%
for $\lambda _{1}>\lambda _{3}$ (with $k_{f}=\left(
k_{f}^{+}+k_{f}^{-}\right) /2$ independent of $\theta $) and $\sigma
_{yx}^{z}=-\frac{9e}{8\pi }$ for $\lambda _{1}<\lambda _{3}.$ The parameters
$c_{x}$ and $c_{y}$ can be calculated numerically, and the result is plotted
in Fig. 4 .

Unlike the 2D Rashba system in which the intrinsic spin Hall conductivity
can be suppressed by the vertex correction completely \cite%
{Inoue04prb,Shen05prl}, the spin Hall conductivity in the present system can
survive in the clean limit. In the case of $\lambda _{1}>\lambda _{3}$ the
vertex correction almost cancels the intrinsic part when the system deviates
from the transition point, $\sigma _{yx}^{z}\approx +$ $0.5\frac{e}{8\pi }$
but has a large residue near the transition point. In the case of $\lambda
_{1}<\lambda _{3}$ the vertex correction is zero, which is consistent with
previous calculation for the cubic Rashba system \cite{Bernevig05prl}.

For a finite density of impurities, numerical results of the total spin Hall
conductivity for different life times are plotted in Fig. 4. The sharp jump
of spin Hall conductivity near the transition point is smeared for the
strong disorder effect. As a result, it is concluded that a jump of the
intrinsic spin Hall conductivity accompanies the change of the Berry phases
near the Fermi surface and it survives after taking into account the
disorder effect of impurities.

\begin{figure}[tbp]
\includegraphics[width=8.5cm]{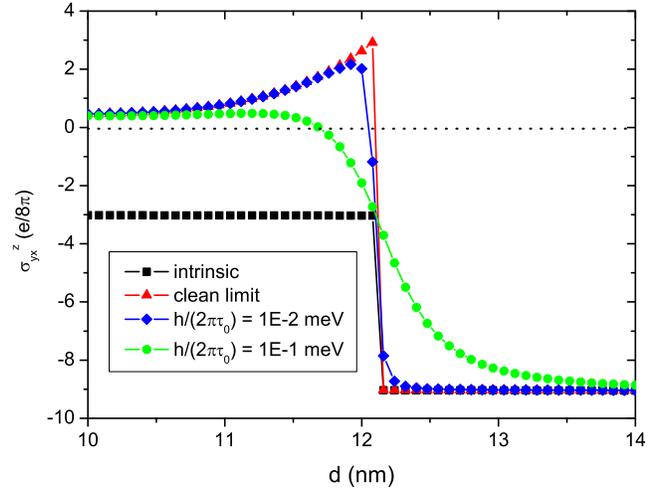}
\caption{{}Variation of spin Hall conductivity $\protect\sigma _{yx}^{z}$
with the thickness $d$ of GaAs quantum well. The material parameters are
given in the text and a given Fermi energy $E_{f}$ is equal to $2.5$ meV.
The squares (black) correspond to the intrinsic part of spin Hall
conductivity; the triangle (red) to spin Hall conductivity in the clean
limit; the diamonds(blue) to $\hbar /\protect\tau _{0}=10^{-2}$ meV; the
circles (green) to $\hbar /\protect\tau _{0}=10^{-1}$ meV. Here $\hbar /%
\protect\tau _{0}=mn_{i}V_{0}^{2}/\hbar ^{2}$.}
\end{figure}

\section{Discussion and summary}

From the calculation above, we established a relation between the
topological quantum phase transition and spin-resolved quantum transverse
transport in the system. The spin Hall effect has been observed
experimentally in both $p$- and $n$-doped semiconductor systems \cite%
{Kato04Science,Wunderlich05PRL} and metals such as aluminum \cite%
{Valenzuela06nature} and platinum \cite{Kimura07prl}. Especially, the
technique of Wunderlich \textit{et al} \cite{Wunderlich05PRL} can be applied
to observe this topological quantum phase transition explicitly. The 2D
hole-doped layer of (Al, Ga)As/GaAs heterojunction is designed as a part of
a $p$-$n$ junction light-emitting diode with a specially designed coplanar
geometry which allows an angle-resolved polarization detection at opposite
edges of the 2D hole system. When an electric field is applied across the
hole channel, a nonzero out-of-plane component of the angular momentum can
be detected whose magnitude depends on the thickness of the heterojunction
for 2D holes. A series of samples with different thickness around $d_{c2}$
are required to detect the jump near the transition point. On the other
hand, as mentioned above, we can also vary Rashba coupling $\alpha $ near
the critical $\alpha _{c}$ by adjusting the gate voltage, and detect the
jump of spin accumulation at edges of the 2D hole quantum well with fixed
thickness to reveal the transition of Berry phase. Technically it is
believed that there is no any obstacle to observe this transition. In short
the topological quantum phase can be characterized by the Berry phase
accumulated by the adiabatic motion of particles on the occupied Bloch
states of hole (or electron). The conventional phase transition is
characteristic of discontinuity of the derivative of the free energy with
respect to temperature. Instead, this novel type of topological quantum
phase transition is revealed by the discontinuity or anomaly of quantum spin
transverse transport in the system.

\acknowledgments This work was supported by the Research Grant Council of
Hong Kong under Grant No.: HKU 7042/06P, and the CRCG of the University of
Hong Kong.


\begin{thebibliography}{99}
\bibitem{Hall-effect} \Editor{Prange R. E. and Girvin S. M.}
\Book{The
Quantum Hall Effect} \Publ{Springer-Verlag, New York} \Year{1987}.

\bibitem{Thouless82prl}
\Name{Thouless D. J., Kohmoto M., Nightingale M. P.
\and den Nijs M.} \REVIEW{Phys. Rev. Lett.}{49}{1982}{405}.

\bibitem{Jungwirth02PRL} \Name{Jungwirth T., Niu Q. \and MacDonald A. H.}
\REVIEW{Phys.
Rev. Lett.}{88}{2002}{207208}; \Name{Onada M. \and Nagaosa N.}
\REVIEW{J.
Phys. Soc. Jpn.}{71}{2002}{19};
\Name{Fang Z., Nagaosa
N., Takahashi K. S., Asamitsu A., Mathieu R., Ogasawara T., Yamada
H., Kamasaki M., Tokura Y. \and Terakura K.}\REVIEW{
Science}{302}{2003}{92}.

\bibitem{Haldane04prl} \Name{Haldane F. D. M.}
\REVIEW{Phys. Rev.
Lett}{93}{2004}{206602}.

\bibitem{Shen04PRB} \Name{Shen S. Q.}
\REVIEW{Phys. Rev.
B}{70}{2004}{081311(R)}.

\bibitem{Chang05prb} \Name{Chang M. C.}
\REVIEW{Phys. Rev.
B}{71}{2005}{085315}; \Name{Chen T.-W., Huang C.-M. \and Guo G. Y.}
\REVIEW{Phys. Rev.
B}{73}{2006}{235309}.

\bibitem{Sheng06prl}
\Name{Sheng D. N., Weng Z. Y., Sheng L. \and
Haldane F. D. M.} \REVIEW{Phys. Rev. Lett.}{97}{2006}{036808}.

\bibitem{Bernevig06SCI} \Name{Bernevig B. A., Hughes T. L. \and Zhang S.-C.} %
\REVIEW{Science}{314}{2006}{1757}.

\bibitem{Berry84} \Name{Berry M. V.}
\REVIEW{Proc. R. Soc. London, Ser.
A}{392}{1984}{45}.

\bibitem{BPZ84Book} \Editor{Meier F. \and Zakharchenya B.
P.} \Book{Optical Orientation} \Publ{North-Holland, Amsterdam} \Year{1984}.

\bibitem{Bulaev05PRL} \Name{Bulaev D. V. \and Loss D.}
\REVIEW{Phys. Rev.
Lett.}{95}{2005}{076805}.

\bibitem{Luttinger56pr} \Name{Luttinger J. M.}
\REVIEW{Phys.
Rev.}{102}{1956}{1030}.

\bibitem{Dresselhaus55} \Name{Dresselhaus G.}
\REVIEW{Phys.
Rev.}{100}{1955}{580}.

\bibitem{Rashba60} \Name{Rashba E. I.}
\REVIEW{Fiz. Tverd. Tela (Leningrad)}{2}{1960}{1224} [\REVIEW{Sov.
Phys. Solid State}{2}{1960}{1109}].

\bibitem{Winkler00prb} \Name{Winkler R.}
\REVIEW{Phys. Rev.
B}{62}{2000}{4245}; \Name{Winkler R.,
 Noh H., Tutuc E. \and Shayegan M.} \REVIEW{Phys. Rev.
B}{65}{2002}{155303}.

\bibitem{Winkler03Book} \Name{Winkler R.}
\Book{Spin-Orbit Coupling Effects in
Two-Dimensional Electron and Hole Systems} \Publ{Springer-Verlag,
Berlin} \Year{2003}.

\bibitem{Zhu94PRB} \Name{Zhu B. F. \and Chang Y. C.}
\REVIEW{Phys. Rev.
B}{50}{1994}{11932}; \Name{Shen S. Q. \and Wang Z. D.}
\REVIEW{Phys. Rev.
B}{61}{2000}{9532}; \Name{Foreman B. A.}
\REVIEW{Phys. Rev.
Lett.}{84}{2000}{2505}.

\bibitem{Miller03PRL}
\Name{Miller J. B., Zumb\"{u}hl D. M., Marcus C. M.,
Lyanda-Geller Y. B., Goldhaber-Gordon D., Campman K. \and Gossard
A. C.} \REVIEW{Phys. Rev. Lett.}{90}{2003}{076807};
\Name{de
Andrada e Silva E. A., La Rocca G. C. \and Bassani F.}
\REVIEW{Phys. Rev.
B}{50}{1994}{8523}; \Name{Cardona M.,
Christensen N. E. \and Fasol G.} \REVIEW{Phys. Rev.
B}{38}{1988}{1806}.

\bibitem{Schliemann05PRB} \Name{Schliemann J. \and Loss D.}
\REVIEW{Phys. Rev.
B}{71}{2005}{085308}.

\bibitem{Note} $\mathbf{A}_{\mu }$ has the so-called "gauge choice". The
gauge we choose is consistent with the following approach: firstly, we add a
term $h\sigma _{z}$ to the Hamiltonian (\ref{HH}) to lift the degeneracy at $%
\mathbf{k}=0$, then calculate the Berry phase through the surface integral
of the gauge-invariant\ Berry curvature and take the limit of $h\rightarrow
0 $ at the last step.

\bibitem{Murakami03sci} \Name{Murakami S., Nagaosa N.  \and Zhang S.-C.} %
\REVIEW{Science}{301}{2003}{1348};
\Name{Sinova J., Culcer D., Niu
Q., Sinitsyn N. A., Jungwirth T. \and MacDonald A. H.}
\REVIEW{Phys. Rev.
Lett.}{92}{2004}{126603};
\Name{Shen S. Q., Ma
M., Xie X. C. \and Zhang F. C.} \REVIEW{Phys. Rev.
Lett.}{92}{2004}{256603}.

\bibitem{Sinova06ssc}
\Name{Sinova J., Murakami S., Shen S. Q.
\and Choi M. S.} \REVIEW{Solid State Commun.}{138}{2006}{214}.

\bibitem{Inoue04prb} \Name{Inoue J. I., Bauer G. E. W. \and Molenkamp L. W.}
\REVIEW{Phys.
Rev. B}{70}{2004}{041303(R)}; \Name{Khaetskii A.}
\REVIEW{Phys.
Rev. Lett.}{96}{2006}{056602}.

\bibitem{Abrikosov63Book}
\Name{Abrikosov A. A., Gorkov L. P. \and
Dzyaloshinski I. E.}
\Book{Methods of Quantum Field Theory in
Statistical Physics} \Publ{Dover, New York} \Year{1963}.

\bibitem{Shen05prl} \Name{Shen S. Q.}
\REVIEW{Phys. Rev.
Lett.}{95}{2005}{187203}; \Name{Zhou B.
Ren L. \and Shen S. Q.} \REVIEW{Phys. Rev. B}{73}{2006}{165303}.

\bibitem{Bernevig05prl} \Name{Bernevig B. A. \and Zhang S. C.}
\REVIEW{Phys. Rev.
Lett.}{95}{2005}{016801}.

\bibitem{Kato04Science}
\Name{Kato Y. K., Myers R. C., Gossard A. C. \and
Awschalom D. D.} \REVIEW{Science}{306}{2004}{1910}.

\bibitem{Wunderlich05PRL}
\Name{Wunderlich J., Kaestner B., Sinova J. \and
Jungwirth T.} \REVIEW{Phys. Rev. Lett}{94}{2005}{047204}.

\bibitem{Valenzuela06nature} \Name{Valenzuela S. O. \and Tinkham M.}
\REVIEW{Nature
(London)}{442}{2006}{176}.

\bibitem{Kimura07prl}
\Name{Kimura T., Otani Y., Sato T., Takahashi S. \and
Maekawa S.} \REVIEW{Phys. Rev. Lett.}{98}{2007}{156601}.
\end{thebibliography}
\end{document}